\documentclass[lettersize,journal]{IEEEtran}
\usepackage{amsmath,amsfonts}
\usepackage{algorithmic}
\usepackage{algorithm}
\usepackage{array}
\usepackage[caption=false,font=normalsize,labelfont=sf,textfont=sf]{subfig}
\usepackage{textcomp}
\usepackage{stfloats}
\usepackage{url}
\usepackage{verbatim}
\usepackage{graphicx}
\usepackage{cite}
\usepackage{xcolor}
\usepackage{siunitx, xfrac}
\usepackage{glossaries}
\usepackage{textpos}
\newcommand*\mygls[1]{%
  \protect\ifglsused{#1}{%
    \glsentryshort{#1}%
  }{%
    \glsentrylong{#1}%
  }%
}

\newacronym{XR}{XR}{Extended Reality}
\newacronym{HR2LLC}{HR2LLC}{High-Rate and High-Reliability Low-Latency Communications}
\newacronym[plural=APs,longplural=Access Points]{AP}{AP}{Access Point}
\newacronym{HMD}{HMD}{Head-Mounted Display}
\newacronym{QoE}{QoE}{Quality of Experience}
\newacronym{FoV}{FoV}{Field of View}
\newacronym{MAC}{MAC}{Medium Access Control}
\newacronym{LoS}{LoS}{Line-of-Sight}
\newacronym{SotA}{SotA}{State of the Art}
\newacronym{mmWave}{mmWave}{Millimeter-Wave}
\newacronym{IMU}{IMU}{Internal Measurement Unit}
\newacronym{ISAC}{ISAC}{Integrated Sensing and Communication}
\newacronym{THz}{THz}{Terahertz}
\newacronym{AoA}{AoA}{Angle of Arrival}
\newacronym{AoD}{AoD}{Angle of Departure}
\newacronym{BHI}{BHI}{Beacon Header Interval}
\newacronym{RIS}{RIS}{Reconfigurable Intelligent Surface}
\newacronym{6DoF}{6DoF}{Six Degrees of Freedom}
\newacronym{http}{HTTP}{hypertext transfer protocol}
\newacronym{lldash}{LL-DASH}{Low-Latency Dynamic Adaptive Streaming Over HTTP}
\newacronym{dash}{DASH}{dynamic adaptive streaming over \mygls{http}}
\newacronym{webrtc}{WebRTC}{Web Real-Time Communication}
\newacronym{ietf}{IETF}{Internet Engineering Task Force}
\newacronym{moq}{MOQ}{Media over QUIC}
\newacronym{2d}{2D}{two-dimensional}
\newacronym{VMAF}{VMAF}{Video Multimethod Assessment Fusion}
\newacronym{MIMO}{MIMO}{Multiple-Input Multiple-Output}
\newacronym{MU-MIMO}{MU-MIMO}{Multi-User MIMO}
\newacronym{MCS}{MCS}{Modulation and Coding Scheme}
\newacronym{NR}{NR}{New Radio}

\begin{document}

\title{Toward Interactive Multi-User Extended Reality using Millimeter-Wave Networking}

\author{Jakob Struye,~\IEEEmembership{Graduate Student Member}, Sam Van Damme, Nabeel Nisar Bhat,~\IEEEmembership{Graduate Student Member}, Arno Troch, Barend van Liempd, Hany Assasa, Filip Lemic, Jeroen Famaey,~\IEEEmembership{Senior Member}, Maria Torres Vega,~\IEEEmembership{Senior Member}

\thanks{J. Struye, N. Nisar Bhat, A. Troch and J. Famaey are with the University of Antwerp and imec, Belgium. S. Van Damme is with Ghent University-imec and KU Leuven, Belgium. B. van Liempd and H. Assasa are with Pharrowtech, Belgium. F. Lemic is with i2Cat Foundation, Spain. M. Torres Vega is with Ghent University-imec and KU Leuven, Belgium.}

        % <-this % stops a space
%\thanks{This paper was produced by the IEEE Publication Technology Group. They are in Piscataway, NJ.}% <-this % stops a space
%\thanks{Manuscript received April 19, 2021; revised August 16, 2021.}
}



\maketitle

\begin{abstract}
Extended Reality (XR) enables a plethora of novel interactive shared experiences. Ideally, users are allowed to roam around freely, while audiovisual content is delivered wirelessly to their Head-Mounted Displays (HMDs). Therefore, truly immersive experiences will require massive amounts of data, in the range of tens of gigabits per second, to be delivered reliably at extremely low latencies. We identify Millimeter-Wave (mmWave) communications, at frequencies between 24 and \SI{300}{\giga\hertz}, as a key enabler for such experiences. In this article, we show how the mmWave state of the art does not yet achieve sufficient performance, and identify several key active research directions expected to eventually pave the way for extremely-high-quality mmWave-enabled interactive multi-user XR.
\end{abstract}

\begin{textblock}{180}(0,8.3)
    \begin{tiny}
    \copyright\ 2024 IEEE. Personal use of this material is permitted. Permission from IEEE must be obtained for all other uses,\\
    \vspace{-6mm}\\
    in any current or future media, including reprinting/republishing this material for advertising or promotional purposes,\\
    \vspace{-6mm}\\
    creating new collective works, for resale or redistribution to servers or lists, or reuse of any copyrighted component\\
    \vspace{-6mm}\\
    of this work in other works.
    \end{tiny}
    \end{textblock}
\section{Introduction}
Since the inception of modern \gls{XR} (which comprises Augmented, Virtual, and Mixed Reality, or AR/VR/MR), \glspl{HMD} have evolved from experimental, bulky, low-resolution devices to sleek, lightweight user-oriented peripherals. More and more applications of VR, where the user is transported to a fully artificial world, as well as its sibling technologies AR and MR, where virtual elements are overlaid onto, or integrated into the physical world, are being widely deployed. \gls{XR} applications include employee training and education, sightseeing tours, and entertainment. We expect the recent release of the Apple Vision Pro MR \gls{HMD} to lead to further mainstream acceptance and adoption of these technologies.

Traditionally, \glspl{HMD} are connected by a wire to a powerful computer, which generates and renders the \gls{XR} content. However, this tether inhibits the users' mobility and immersion, and can result in a tripping hazard. As an alternative, recent \glspl{HMD} offer on-board processing capabilities. These are often focused on AR/MR, which does not require generating a full \SI{360}{\degree} environment (e.g., Apple Vision Pro). For VR, they are limited to rendering lower-quality content due to their constrained computational capabilities (e.g., Meta Quest 3). The obvious solution is to provide a high-data-rate wireless connection between the \gls{HMD} and rendering location (e.g., a nearby computer or (edge) cloud server)~\cite{mmWaveVR, cellularVR}. Several \gls{SotA} VR \glspl{HMD}, such as the Meta Quest 3, offer wireless connectivity using Wi-Fi on the \SI{5}{\giga\hertz} frequency band. However, as uncompressed \gls{XR} content may require tens of gigabits per second, a high compression rate is applied, resulting in increased latency and visual artifacts. Achieving high \gls{QoE} in \gls{XR} is further complicated by the fact that it requires a \textit{motion-to-photon latency} of at most \SI{20}{\milli\second}, to avoid cybersickness~\cite{mmWaveVR}. This encompasses the total latency between a user's motion and the corresponding update of the visual image on the \gls{HMD}. These requirements become even more stringent in interactive multi-user \gls{XR} experiences, where users interact with each other, as well as with the virtual or hybrid environment. They may be co-located, or may participate from different physical locations. Enabling such seamless interactivity requires extremely low latency, alongside dense multi-user and wide-area connectivity.

\gls{XR} thus requires a combination of high data rate, high reliability, and low latency network connectivity, known as \gls{HR2LLC}~\cite{hr2llc}. Due to the limited bandwidth and high congestion of the sub-\SI{6}{\giga\hertz} frequency bands, \gls{mmWave} wireless communications (i.e., 24--\SI{300}{\giga\hertz}) has been identified as a prime enabler of wireless \gls{XR}~\cite{mmWaveVR}.  The multi-gigahertz bands available in \gls{mmWave} offer data rates of up to tens of gigabits at extremely low latency, but pose their own set of challenges to achieving consistent transmission quality. Notably, \gls{mmWave} experiences high path and penetration loss. This hinders the establishment of consistently high-gain links and renders them prone to blockage, including by users themselves. Ensuring \gls{HR2LLC} at \gls{mmWave} frequencies requires a combination of large antenna arrays, directional beamforming, and multi-\gls{AP} connectivity. This is especially challenging in multi-user interactive \gls{XR}, featuring highly mobile users within a confined space.

In this article, we present our vision for future truly high-\gls{QoE} interactive multi-user \gls{XR} experiences. We present an overview of the scenario and the \gls{HR2LLC} requirements it poses on the network. We expect that \gls{mmWave} wireless communications will be an essential building block towards consistently fulfilling these requirements. We identify the shortcomings of the current \gls{mmWave} \gls{SotA}, and explore potential avenues towards addressing these, including \gls{XR}-specific beamforming, Reflective Intelligent Surfaces, channel access optimizations, Integrated Sensing and Communication, low-latency real-time streaming protocols and multi-user human-centric perception for performance evaluations.

\section{Collaborative Wireless XR}
Arguably the most technically challenging form of \gls{XR} is collaborative wireless \gls{XR}. The inherent difficulty of achieving \gls{HR2LLC} wirelessly is further amplified by the interactivity of such experiences. Common solutions such as content caching and heavy-duty compression become ineffective when content generation is dependent on several users' real-time actions~\cite{caching}. Furthermore, wireless resource allocation when several users require an uninterrupted high-quality link is extremely challenging. We provide further details of the physical scenario, discuss the requirements, and summarize current solutions and their shortcomings.

\begin{figure*}[!t]
\centering
\includegraphics[width=\textwidth]{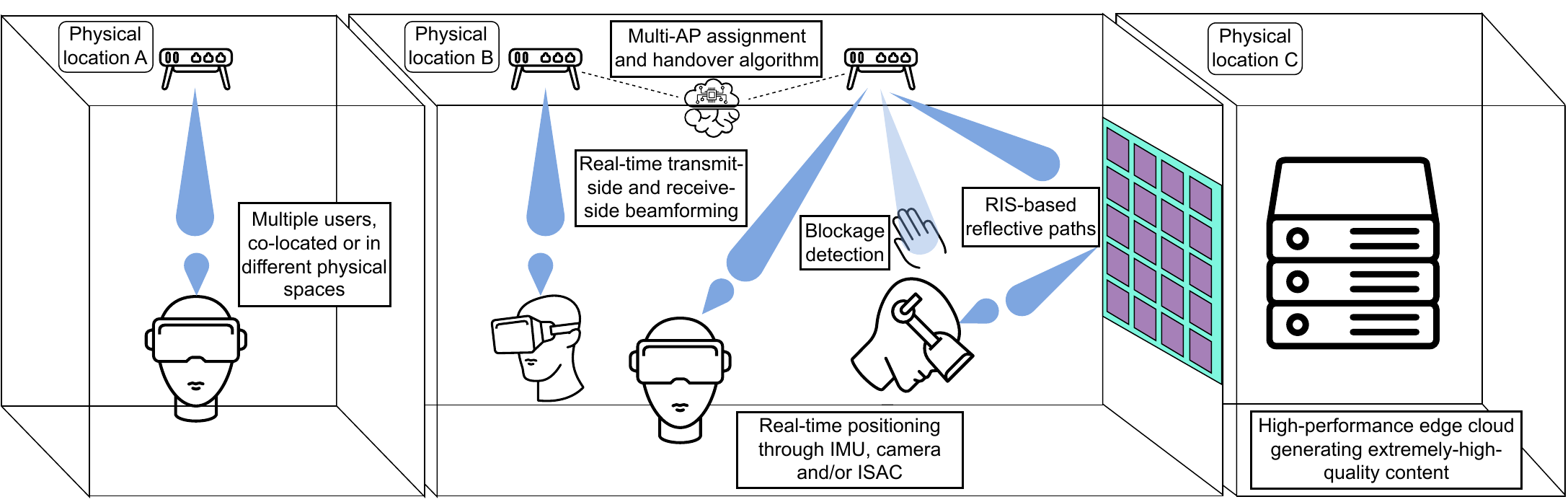}
\caption{Overview of the envisioned multi-user interactive \gls{XR} scenario.}
\label{fig:overview}
\end{figure*}

\subsection{System Architecture}
Collaborative \gls{XR} encompasses multi-user, interactive experiences, where users roam freely within a shared virtual (or mixed virtual-physical) environment, as shown in Fig.~\ref{fig:overview}. Users may be separated across multiple physical locations (A and B in the figure), with each location hosting one or more users. Geographically separated users are projected into each other's spaces. Content is generated in real-time, likely at another physical location. In the figure, content is computer-generated in an edge cloud at physical location C. To facilitate realistic and unconstrained \gls{6DoF} motion, \glspl{HMD} estimate their own pose through \glspl{IMU} and/or built-in cameras. Furthermore, they incorporate a \gls{mmWave} antenna, and content is delivered by one or more \gls{mmWave} \glspl{AP}, mounted on walls or the ceiling. \Glspl{HMD} and \glspl{AP} beamform towards each other, illustrated in blue in the figure. In multi-\gls{AP} deployments, a centralized algorithm orchestrates \gls{AP} assignment and handovers, with its objective commonly being to maximize the lowest \gls{QoE} among all users. Crucially, when signal degradation indicates upcoming blockage, the algorithm should avoid quality degradation by reassigning the affected user. Finally, the figure incorporates several experimental systems for further improving the \gls{QoE}, further covered in Sec.~\ref{sec:solutions}, including novel approaches to real-time beamforming, establishing viable reflected links, and pose estimation.

\subsection{Requirements}
\gls{HR2LLC} summarizes the network requirements of collaborative wireless \gls{XR}~\cite{hr2llc}. The \textbf{high rate} is determined by the quality of the content, which is in turn limited by the hardware specifications of the \gls{HMD}~\cite{hardware}. Without compression, the Meta Quest 3 requires between 15.75 and \SI{26.25}{Gbps} depending on the refresh rate. 
% 2064 x 2208 pixels per screen, 2 screens, 72-120hz, 24 bits per pixel
While compression may reduce this staggering requirement, this introduces an additional (de)compression delay. This may impact the \textbf{low latency} requirements, with the \textit{motion-to-photon} latency limit for XR being commonly defined as \SI{20}{\milli\second}, meaning the result of any user motion must be reflected in-experience within \SI{20}{\milli\second} to avoid nausea~\cite{mmWaveVR}. Depending on other factors, this may leave between 5 and \SI{8}{\milli\second} for one-way network transmissions. Any content not arriving on time is essentially lost, which is highly impactful given the \textbf{high-reliability} requirement. This reliability requirement is defined at two levels. The \textit{intra-image reliability} determines the fraction of an image that needs to arrive on time in order to be considered as complete. The exact target fraction depends on redundancy in any compression algorithm, along any reconstruction algorithms extrapolating missing data from arrived content. The \textit{inter-image reliability} defines how many images may be lost without unacceptable impact on \gls{QoE}, both overall and within a single loss burst. The exact threshold depends on the specific experience, but may reach as high as \SI{99.999}{\percent}, or roughly 1 missed image per 15 minutes. While there are many methods for measuring the \gls{QoE} of interactive \gls{XR}~\cite{qoe}, fulfilling the \gls{HR2LLC} requirements is always a necessary condition for achieving satisfactory results.

\subsection{Existing Solutions and Shortcomings}

Achieving \gls{HR2LLC} wirelessly will require \gls{mmWave} communications. Overcoming \gls{mmWave}'s high path loss and susceptibility to blockage demands a specialized approach. Through antenna arrays consisting of many elements, along with \textit{beamforming}, a process in which energy is focused in a carefully selected direction, a sufficiently high gain can be achieved even with modest overall energy budgets. As the range of directions in which a \gls{mmWave} antenna can beamform is limited, they effectively have a limited \gls{FoV}, usually between 60 and \SI{120}{\degree}. As such, a connection can easily be interrupted by the user walking around or even simply turning their head. Enabling consistent high-gain coverage in the face of blockage and limited \glspl{FoV} therefore requires multiple, spatially separated \glspl{AP}.

At the protocol level, both 5G and Wi-Fi offer explicit \gls{mmWave} support within their specifications. 5G \gls{NR} supports several bands within the \gls{mmWave} range, with most deployments using licensed bands between 24.5 and \SI{29.5}{\giga\hertz}. Wi-Fi, in turn, supports \gls{mmWave} using unlicensed bands between 57 and \SI{70}{\giga\hertz}. Frequencies above \SI{100}{\giga\hertz} are being explored for potential use in 6G. Concerning current-day protocols, both 5G and Wi-Fi consider \gls{XR} as an explicit use case for their \gls{mmWave} functionality in official documentation. However, 5G defines this as an outdoors use case with expected data rates in the tens of megabits per second, while Wi-Fi describes indoors experiences with multi-gigabit requirements. As the latter overlaps closely with the scenario described in this article, we adopt Wi-Fi's terminology (e.g., ``\gls{AP}'') throughout the article, but note that all solutions described in Sec.~\ref{sec:solutions} are protocol-agnostic.

\begin{figure}[!t]
\centering
\includegraphics[width=\linewidth]{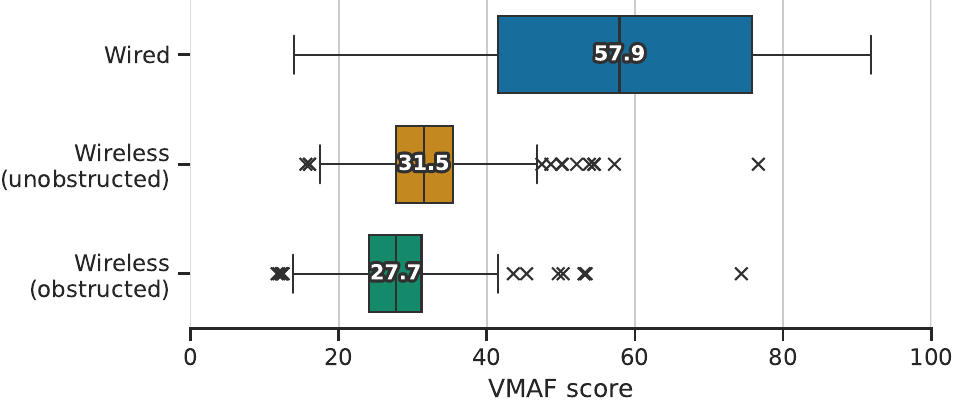}
\caption{Perceived video quality comparison for the HTC Vive in wired and (un)obstructed wireless scenarios. A higher VMAF score is better.}
\label{fig_Vive}
\end{figure}

Some \gls{mmWave} solutions for wireless \gls{XR} have been brought to market. Most notably, the HTC VIVE Wireless Adapter replaces the usual cable with a wireless \gls{mmWave} bridge, running a custom protocol. To understand the performance of these wireless options, we compared the performance in wired and wireless mode, additionally considering the scenario in which the wireless link is obstructed. We transmitted the same content, computer-generated in real-time, in the three scenarios, calculating the \gls{VMAF} score to assess video quality objectively. The \gls{VMAF} score (0-100) is a per-frame quality similarity metric compared to a reference recording of the content, with 100 indicating no quality reduction compared to the reference. Fig.~\ref{fig_Vive} shows the mean \gls{VMAF} scores, along with box plots of their per-frame values. Due to frame rate instability, the \gls{VMAF} score of the wired scenario is already suboptimal despite there being no compression or data loss. The \gls{VMAF} drops significantly with a clear wireless link, indicating significant compression occurs as to not saturate the wireless link capacity. Blockage causes an additional reduction in perceived quality.

More recently, the Meta Quest line of \glspl{HMD} offers a wirelessly streamed solution through users' existing \SI{5}{\giga\hertz} Wi-Fi deployments rather than requiring additional hardware. Even with the recently introduced H.264+ compression option, the bitrate is still limited to \SI{400}{Mbps}, leading to readily noticeable compression artifacts, substantiating the need for mmWave.

\begin{figure}[!t]
\centering
\includegraphics[width=\linewidth]{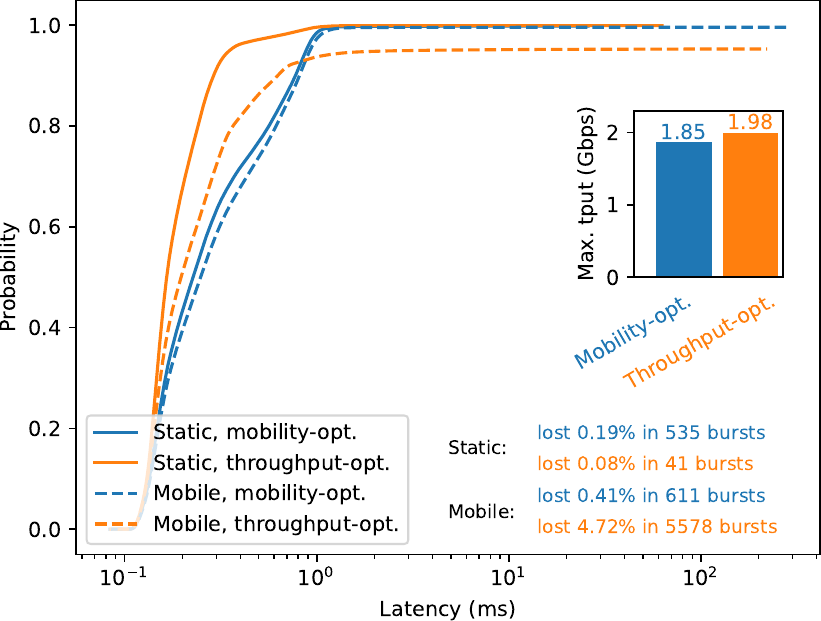}
\caption{Throughput, latency and loss with \gls{SotA} mmWave hardware, for a single \gls{AP} and user, both static and under moderate motion (\SI{45}{\degree\per\second}), with parameters optimized for either throughput or mobility.}
\label{fig_EVK}
\end{figure}

To assess the capabilities of \gls{SotA} \gls{mmWave} hardware, we evaluated the data rate, latency and loss under motion of a mmWave link provided by Pharrowtech. Here we evaluated two sets of \gls{MAC}-level parameters, aimed at optimizing data rate and adaptation to mobility, respectively. This showed that optimizing \gls{MAC}-level parameters boosts throughput, and that moderate motion can lead to an increase in packet loss, even with an unobstructed \gls{LoS} and parameters optimized for mobility. This is illustrated in Fig.~\ref{fig_EVK}, where, in each case, there is room for improvement in reducing loss to meet high-\gls{QoE} \gls{XR} standards. This indicates the need for developing proactive beamforming approaches.
Overall, \gls{SotA} solutions are unable to fulfill the \gls{HR2LLC} requirements of current-day \glspl{HMD}. Future \glspl{HMD} are expected to drastically increase specifications, enabling truly realistic experiences but also further inflating \gls{HR2LLC} requirements, mainly in terms of data rate.

Even experimental \gls{mmWave} Wi-Fi hardware currently fails to achieve the maximal \gls{MCS} supported by the specification, which would offer a data rate of \SI{8.5}{Gbps}. In parallel, two optional features of the specification can improve the data rate orthogonally. Through channel bonding and aggregation, up to four channels may be combined to linearly increase the data rate. This is partially supported by the devices above, and we consider this a viable way to increase effective data rate for interactive \gls{XR}. Next, \gls{MIMO} can increase Wi-Fi data rate by, theoretically, a factor 8, through multiple antenna arrays leveraging multipath propagation. However, to increase data rate towards a single \gls{XR} device, this would require equipping those with multiple (or larger, logically sub-divisible) antenna arrays. Given cost and physical size constraints, we do not consider this to be viable for wearable \gls{XR} devices. Furthermore, based on several measurement campaigns, power consumption of a simple \gls{mmWave} Wi-Fi chip is, at worst, \SI{1}{\watt} higher than with sub-\SI{6}{\giga\hertz} Wi-Fi during active transmission. As overall power consumption of modern \glspl{HMD} is around \SI{10}{\watt}, the reduction in battery life from incorporating \gls{mmWave} is modest. However, \gls{MIMO} would further increase this consumption, with noticeably reduced battery life eventually leading to a reduction in longer-term \gls{QoE}. We do note that \gls{MU-MIMO}, in which the multiple streams are sent to \textit{different} receiving \gls{XR} devices, would only require additional hardware support at the \glspl{AP}, making this a cost-effective way of supporting more users within a physical environment.

Independent of the networking approach, \gls{XR}'s strict latency requirements are often alleviated through Asynchronous Time-Warp (ATW)~\cite{atw}. With this algorithm, images generated based on a stale orientation measurement are perturbed to offset for recent rotational motion, reducing the effective motion-to-photon latency in some use cases. However, this can only address the motion of the viewpoint; other visible physical motions (e.g., in tele-operation) are largely unaffected by ATW, meaning it can not generally reduce effective latency.

\section{Open Challenges and Way Forward}\label{sec:solutions}
Above, we argued that \gls{mmWave} networking is necessary for high-\gls{QoE} interactive multi-user wireless \gls{XR}, but showed that \gls{SotA} performance of \gls{mmWave} solutions does not yet suffice. In this section, we discuss several avenues where \gls{XR}-centric research is ongoing, but more efforts are needed to achieve truly immersive experiences.
\subsection{Beamforming for mobile users}
\begin{figure}[!t]
\centering
\includegraphics[width=\linewidth]{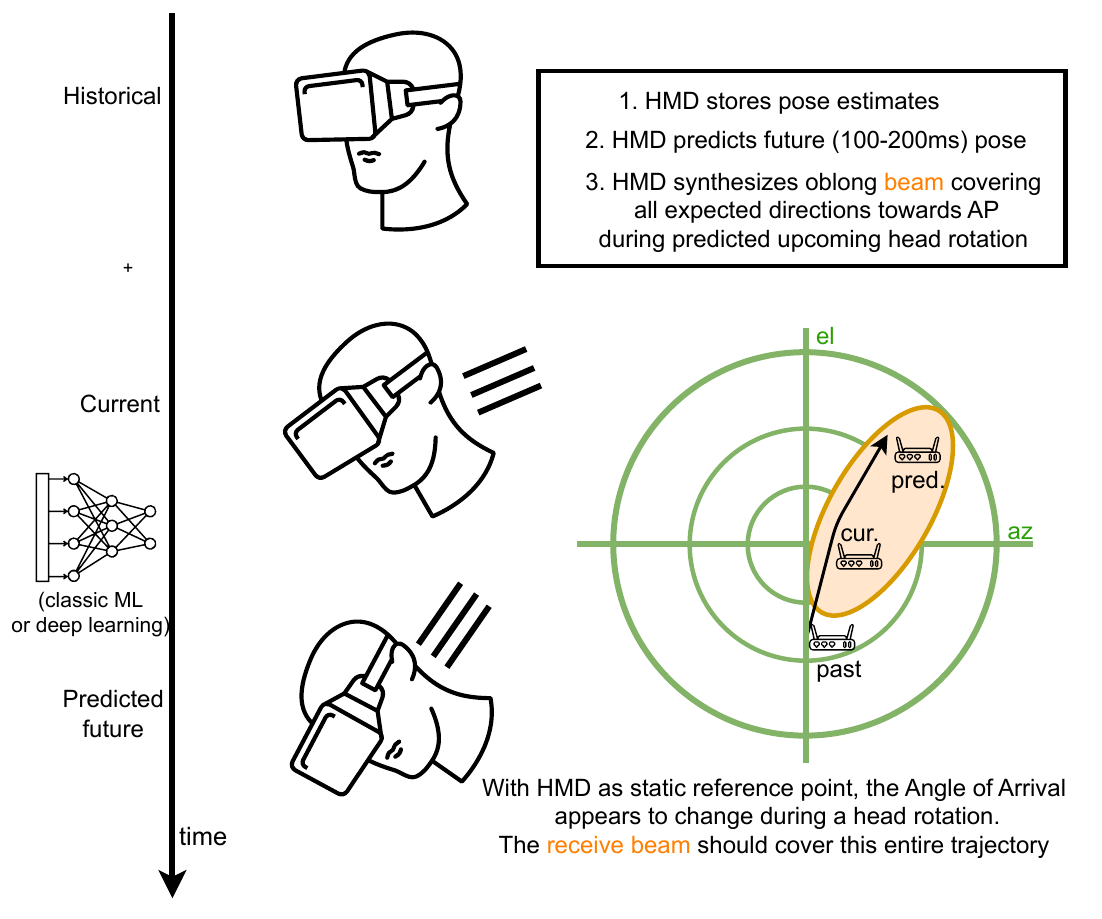}
\caption{\gls{HMD}-side beamforming should proactively adapt to expected upcoming rotations, such that receive gain remains consistently high during rapid user motion.}
\label{fig_covrage}
\end{figure}

In many present-day \gls{mmWave} deployments, transmit-side beamforming suffices to achieve a performant link. The \gls{HR2LLC} requirements of \gls{XR}, however, may necessitate additional receive-side beamforming at the \gls{HMD}. Beamforming at this side is inherently more challenging in mobile \gls{XR}; most user rotation changes the \gls{AoA} at the \gls{HMD} drastically, but has minimal impact on the \gls{AoD} at the static \gls{AP}. As such, mobile \gls{XR} necessitates the development of receive-side beamforming approaches which adapt to user rotation with minimal latency, or even proactively. To this end, algorithms can leverage the plethora of sensors already available on modern-day \glspl{HMD}~\cite{pos_bf}. Through \glspl{IMU} and on-device cameras, \glspl{HMD} can accurately estimate their own position and orientation in real-time. By combining this \textit{context information} with the fixed location of \glspl{AP}, the \gls{HMD} can beamform towards the \gls{AP} directly, foregoing the time-consuming beam searching algorithms prevalent in beamforming approaches. In addition, the \gls{HMD} could predict its upcoming motion and form a receive beam that will consistently cover the \gls{AP} during this rotation, as shown in Fig.~\ref{fig_covrage}~\cite{covrage}.

\subsection{Reconfigurable Intelligent Surfaces}
\glspl{RIS} are passive wall-mountable \textit{metasurfaces} whose reflective properties can be altered in real-time~\cite{ris}. By intelligently controlling the reflection angle of incident beams, high-gain reflected paths can be established where these would otherwise not be viable. An intelligent, advanced resource scheduler could take these into account to maximize the \gls{QoE} in a large \gls{XR} deployment. As the passive elements that comprise a \gls{RIS} are low-complexity and therefore low-cost, \gls{RIS} is a promising avenue towards increasing coverage in \gls{XR} scenarios at a low cost. Existing prototypes at both \gls{mmWave} and lower frequencies have proven the viability of the concept. However, practical and experimental works on \gls{mmWave} \gls{RIS} are still scarce, and none consider the extremely high data rates required for \gls{XR}. Prototypes convincingly demonstrating the viability of \gls{RIS}-enhanced interactive \gls{XR} would aid in paving the way towards affordable consumer-grade solutions. 

Furthermore, we identify a need for \textbf{real-time} \gls{RIS} configuration algorithms. \gls{SotA} configuration algorithms are already well-performing, but rely on high-complexity optimizers, leading to runtimes of minutes to hours~\cite{ris_performance}. These are focused on fully static environments, where a single iteration of such algorithms leads to a permanent configuration. However, deploying \glspl{RIS} in interactive \gls{XR} scenarios necessitates heuristic approaches with runtimes in the range of milliseconds. We identify Deep Reinforcement Learning as a potential enabler to this end.

\subsection{High-Data Rate Low-Latency Channel Access}
\begin{figure}[!t]
\centering
\includegraphics[width=\linewidth]{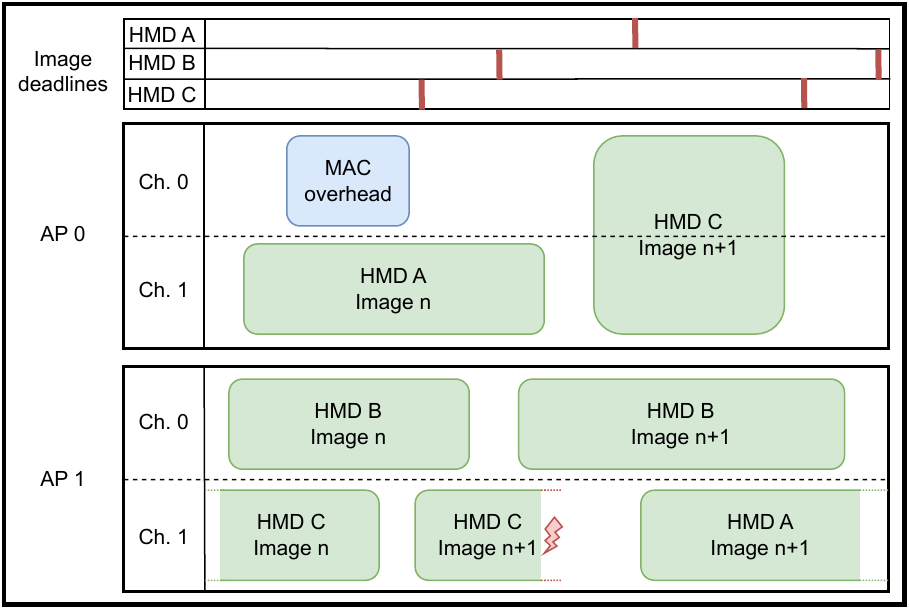}
\caption{Adaptive scheduling for a multi-AP, multi-user, multi-channel deployment. During transmission of image $n+1$ to \gls{HMD} C, the connection is suddenly interrupted (e.g., due to hand blockage), after which \gls{HMD} C is rapidly moved to another \gls{AP} and given two channels to ensure the image deadline is met. To facilitate this, \gls{HMD} A switches to the other \gls{AP}.}
\label{fig_scheduling}
\end{figure}

Especially in multi-user scenarios, interactive \gls{XR} poses its own challenges and opportunities in terms of scheduling channel access. Most traffic is downstream, meaning orchestration can easily occur in a centralized controller. Even in multi-\gls{AP} deployments, most of the scheduling control overhead can occur over the wired, fully reliable network. The main challenge of \gls{XR} traffic is the strict per-image deadline. An image \textit{must} arrive fully at the \gls{HMD} before the time it is intended to be displayed, or else it is fully lost. This makes efficient interweaving of traffic towards multiple \glspl{HMD} highly challenging. Modern solutions such as channel aggregation and bonding, allowing for dynamic bandwidths, further complicate the resource allocation challenge. In addition, \gls{mmWave} transmission schedules often incorporate repeating periods for \gls{MAC}-layer tasks such as device discovery, association and beamforming, during which no application traffic can be transmitted. For example, the \gls{BHI} for \gls{mmWave} Wi-Fi reoccurs every \SI{100}{\milli\second} and may take several milliseconds each time. Traffic must be scheduled around this carefully, taking deadlines into account~\cite{scheduling}. In addition, an optimal scheduler would be aware of each \gls{HMD}'s refresh cycle, maximizing the percentage of images arriving on time. This requirement could be alleviated somewhat if \glspl{HMD} support a variable refresh rate, with which a screen update could be briefly postponed until an image is fully received. Fig.~\ref{fig_scheduling} shows a schedule taking \gls{MAC}-layer overheads, image deadlines and sudden blockage into account.

\subsection{Integrated Sensing and Communication}
\gls{XR} applications not only require low latency and high-speed communication, but also accurate and real-time sensing of user pose. While the \gls{HMD} pose is needed for rapid beamforming, the continuous and accurate estimation of the \textit{full body pose} is needed for applications where physical actions are translated into the \gls{XR} environment. While current \gls{SotA} solutions for \gls{XR} pose estimation often rely on cameras or hand-held controllers, these approaches come with significant drawbacks. Camera-based methods infringe the privacy of users, require a well-lit environment, and are not scalable. Also, hand-held controllers can be unintuitive to users. Therefore, the overall setup becomes expensive and complex. 

Instead, pose estimation can also occur through the use of wireless signals. In particular, Wi-Fi signals have been used for many \textit{sensing} applications such as pose estimation, gesture recognition, and localization~\cite{liu2019wireless}. This concept of re-using communication signals for sensing is known as \gls{ISAC}. The standout advantage of the sensing approach is its limited additional cost. While Wi-Fi sensing has primarily focused on sub-\SI{6}{\giga\hertz} signals, the constrained bandwidth at these frequencies limits sensing resolution. In contrast, we identify an untapped potential in \gls{mmWave}, where vast bandwidth significantly enhances sensing accuracy. For instance, leveraging a \SI{2}{\giga\hertz} bandwidth at \SI{60}{\giga\hertz} can yield an impressive \SI{15}{\centi\meter} raw resolution in localization applications. Initial results on \gls{mmWave} \gls{ISAC} are promising, however some open challenges remain on the path towards real-time body pose estimation through \gls{mmWave}. Specifically, current estimation performance is highly sensitive to both the body shapes of individuals and the surrounding environment~\cite{liu2019wireless}. As such, we identify the need for more robust estimation algorithms, and expect future Deep Learning-based solutions to fulfill this need.
\subsection{Low Latency End-to-End Streaming}
To ensure consistently high \gls{QoE}, special care must also be taken at the application streaming level, where the system must adapt in real-time to context changes. With a context such as \gls{mmWave}, given its volatile nature, special care is needed to ensure the user's perception of \gls{6DoF} video remains consistent~\cite{survey_xrdelivery}. 
The system must incorporate measurements from client and server to improve the real-time reactiveness of the streaming, such as on-the-fly parameter tuning related to the viewport of the user. Although low-latency video streaming protocols such as \gls{lldash} and \gls{webrtc} are already in used for traditional 2D video, their translation to immersive 3D content, especially when transmitted over \gls{mmWave}, is not straightforward, as these protocols are constrained in terms of processing power, throughput, and latency. Recently, the \gls{ietf} initiated the \gls{moq} working group aiming to develop a simple low-latency media delivery solution addressing use cases including live video streaming, gaming, and media conferencing at scale. Currently, only early results on \gls{moq} are available, and solutions for immersive media use cases are still to be developed~\cite{MOQ2023}. 

\subsection{Multi-User, Human-Centric Perception}
In order to align the aforementioned technologies with human perception, experience and interaction (i.e. human-centric multimedia) and to assess the effectiveness of the above approaches to \gls{QoE} improvement, we need detailed, accurate and expansive evaluations. Traditionally, this human perception evaluation has been performed by means of subjective methodologies, relying on user feedback gathered through questionnaires and prompts. However, these suffer from individual biases, are not scalable and may disrupt a person's experience~\cite{survey_xrdelivery}. Such disruptions impact the evaluation, as breaks in presence and immersion tend to significantly alter the experience. Moreover, subjective evaluations are a posteriori, meaning the full experience is rated at once. Given the volatile nature of \gls{mmWave}, it is fundamental to move towards less intrusive and more real-time assessments of perception. Ideally, objective metrics, driven by physiological data, would provide a more immediate result without requiring conscious assessment. These methodologies are still at their infancy and must thus be investigated and integrated into immersive media experience evaluation. 

Moreover, these assessments are currently performed on the individual, meaning the individual perception and immersion are assessed. However, in multi-user (collaborative) environments, the multiple users share a common environment and goal. Therefore, a focus shift is required from the level of the individual user to the collective perception and experience of the group. This group cohesiveness, or the extent to which group members are attracted to the group and its goals, is affected by a plethora of factors. These include shared cognition (the relationship between group and application/system), shared awareness (the mutual relationship between the group’s individual members), shared engagement (the relationship between group and its common goal) and the individual perception and immersion. The level in which they affect the perception will depend on context factors such as hardware, network, use case, subject and emotional state. As such, this focus shift from individual to collective, multi-user, human-centric perception poses an important and interesting direction for further research.

\section{Conclusion}
In this article, we presented our vision for future deployments of extremely-high-quality interactive multi-user \gls{XR} experiences. Multiple users can roam freely in a shared, collaborative experience, while they may be co-located or in different locations physically. A high-performance edge cloud handles real-time content-generation, while wireless \gls{mmWave} links provide the last-hop connection to \glspl{HMD}. Evaluation of several \gls{SotA} hardware solutions showed that current \gls{mmWave} technology cannot fulfill the extreme requirements of the envisioned scenario. We then identified several key active research tracks towards achieving consistently high-\gls{QoE} interactive multi-user \gls{XR}, discussing ongoing work and open challenges. We are confident that additional effort along these tracks will eventually lead to the realization of this scenario, enabling a plethora of novel applications and experiences.

\section*{Acknowledgments}
This research was partially funded by the ICON project INTERACT and Research Foundation - Flanders (FWO) project WaveVR (Grant number G034322N). INTERACT was realized in collaboration with imec, with project support from VLAIO (Flanders Innovation and Entrepreneurship). Project partners are imec, Rhinox, Pharrowtech, Dekimo and TEO. This work is partially supported by the European Commission through the Horizon Europe JU SNS project Hexa-X-II (Grant Agreement no. 101095759). Nabeel Nisar Bhat and Sam Van Damme are supported by an FWO SB PhD fellowship (Grant numbers 1SH5X24N and 1SB1822N respectively). The work of Filip Lemic was supported by the Spanish Ministry of Economic Affairs and Digital Transformation and the European Union – NextGeneration EU, in the framework of the Recovery Plan, Transformation and Resilience (Call UNICO I+D 5G 2021, ref. number TSI-063000-2021-7); the European Union's Horizon Europe's research and innovation programme under grant agreement nº 101139161 — INSTINCT project; and MCIN / AEI / 10.13039 / 501100011033 / FEDER / UE HoloMit 2.0 project (nr. PID2021-126551OB-C21).

\begin{IEEEbiography}[{\includegraphics[width=1in,height=1.25in,clip,keepaspectratio]{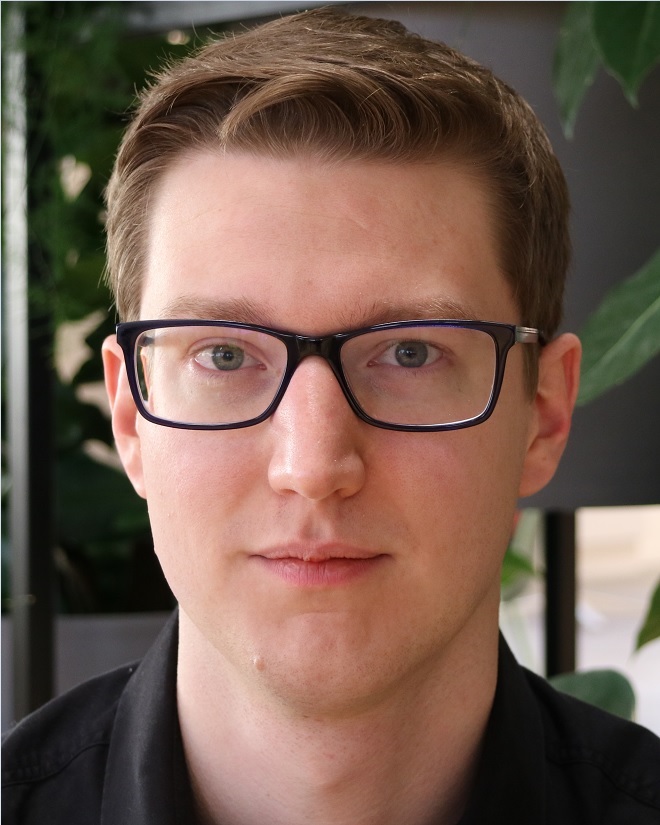}}]{Jakob Struye}
is a Ph.D. researcher in the field of wireless networking at the IDLab research group (University of Antwerp) and imec research institute, Belgium. He obtained his B.Sc (2015) and M.Sc. (2017) in Computer Science at the University of Antwerp. His current research focuses on leveraging extremely-high-frequency wireless networks in the millimeter-wave bands to improve the performance of truly wireless Extended Reality experiences, and has experience in applying Artificial Intelligence to time series prediction problems.
\end{IEEEbiography}

\begin{IEEEbiography}[{\includegraphics[width=1in,height=1.25in,clip,keepaspectratio]{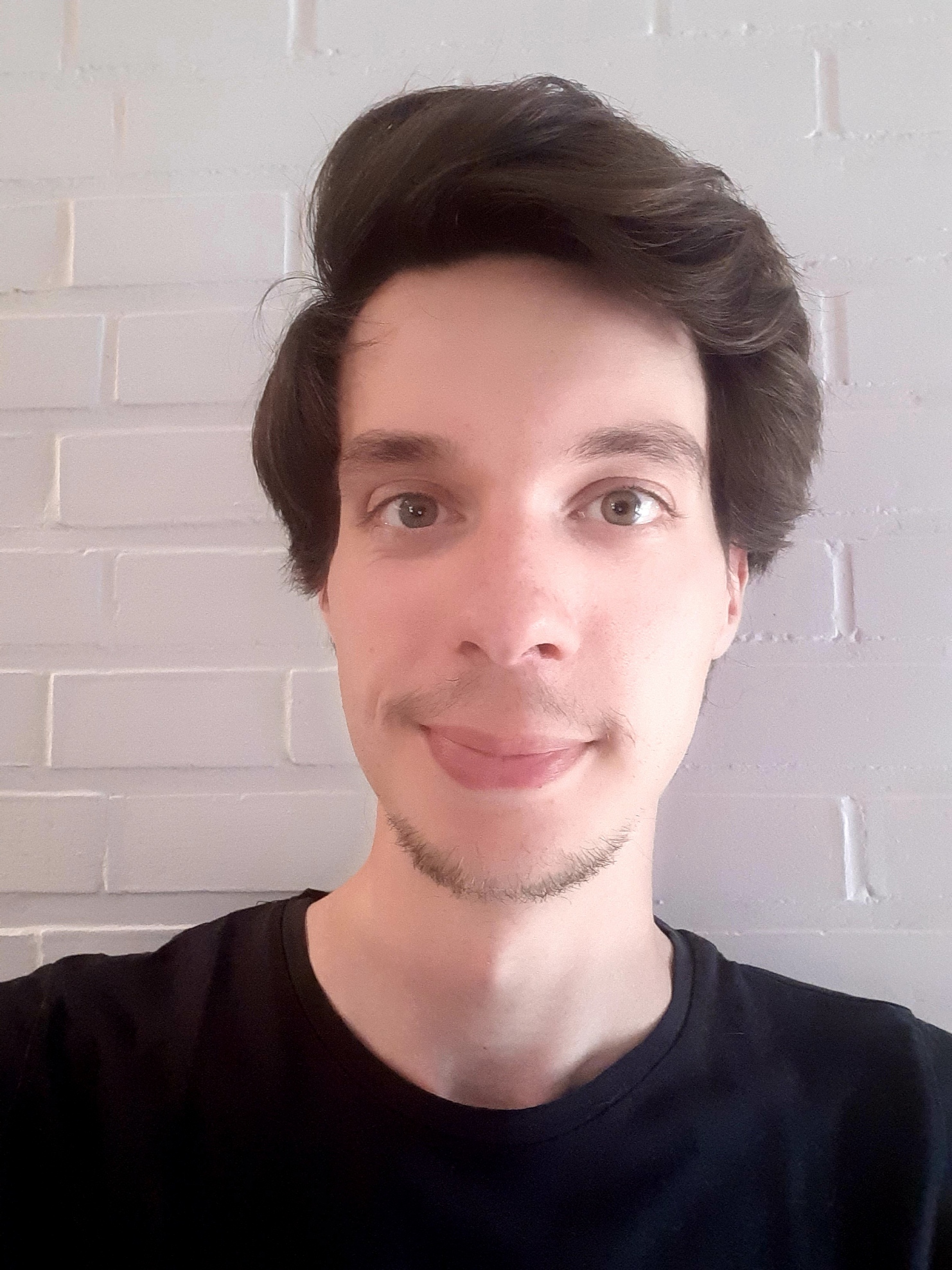}}]{Sam Van Damme}
is a Joint PhD Researcher affiliated with the Internet Technology and Data Science Lab (IDLab) at the Department of Information Technology (INTEC), Ghent University, Belgium and e-Media Research Lab, KU Leuven, Belgium. He obtained his B.Sc. (2016) and M.Sc. (2019) in Computer Science Engineering at Ghent University, Belgium. He has expertise in Quality-of-Experience modelling and assessment, Human-Computer Interaction, and Haptic Feedback with a main focus on immersive multimedia technologies such as Virtual and Augmented Reality.
\end{IEEEbiography}

\begin{IEEEbiography}[{\includegraphics[width=1in,height=1.25in,clip,keepaspectratio]{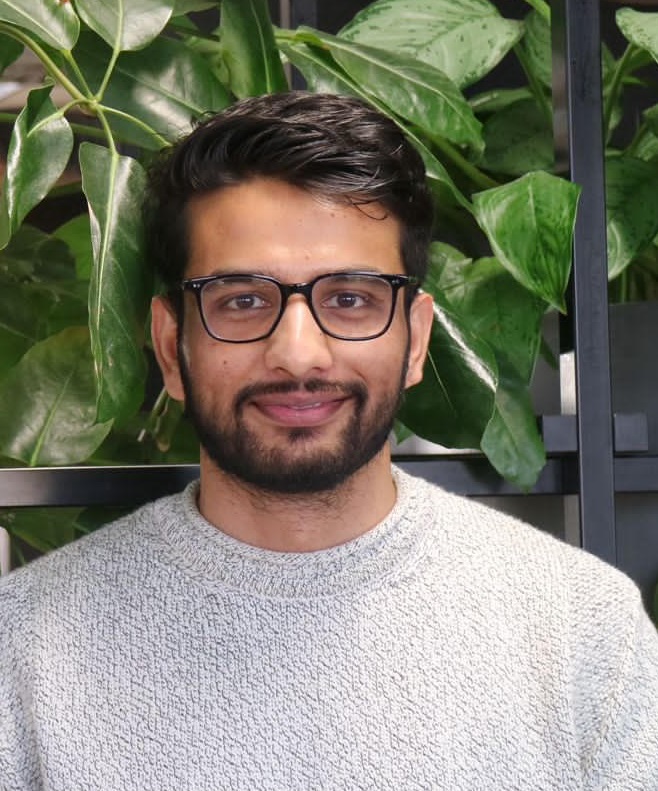}}]{Nabeel Nisar Bhat}
is a Ph.D. researcher in the field of Joint Communication and Sensing at the IDLab research group (University of Antwerp) and imec research institute, Belgium. He obtained his  M.Sc. (2021) in Communications and Computer Networks Engineering at Politecnico di Torino. His current research focuses on leveraging mmWave communication signals for pose estimation in Extended Reality applications. He has experience in signal processing, wireless communications, and deep learning.
\end{IEEEbiography}

\begin{IEEEbiography}[{\includegraphics[width=1in,height=1.25in,clip,keepaspectratio]{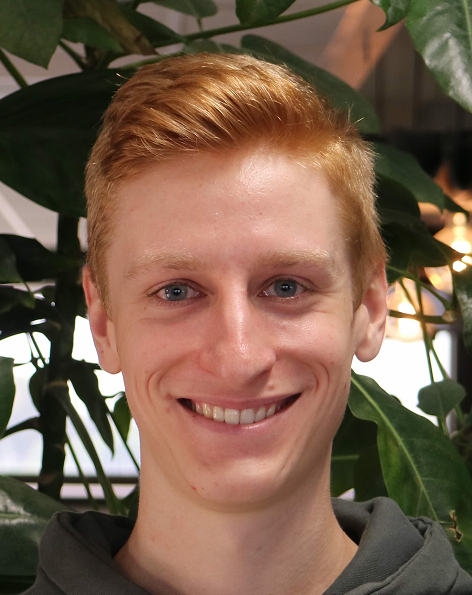}}]{Arno Troch}
is a Ph.D. researcher in the field of wireless and cellular networking at the IDLab research group (University of Antwerp) and imec research institute, Belgium. He obtained his B.Sc. (2021) and M.Sc. (2023) in Computer Science at the University of Antwerp. His current research focuses on mesh networking at millimeter-wave and sub-terahertz frequencies for future cellular networks. His main interests are wireless and cellular networking, network modelling, and vehicular communications.
\end{IEEEbiography}

\begin{IEEEbiography}[{\includegraphics[width=1in,height=1.25in,clip,keepaspectratio]{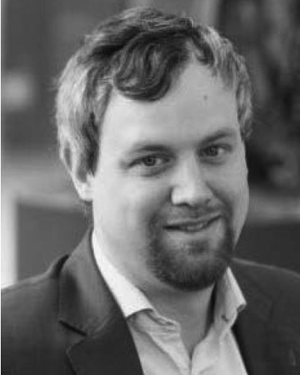}}]{Barend van Liempd}
received the B.Sc. and M.Sc. degrees in EE from Eindhoven University of Technology (2009, 2011), and Ph.D. from Vrije Universiteit Brussel (2017), with a focus on tunable RF front-end circuits. He worked at imec, Belgium (2011-2022), first as R\&D Engineer, and later Senior Researcher, Program Manager and R\&D Manager. Then, he joined Pharrowtech as Manager Hardware Engineering. He published over 60 articles and patents and received the 2015 NXP Prize and 2019 Lewis Winner Award. 
\end{IEEEbiography}

\begin{IEEEbiography}[{\includegraphics[width=1in,height=1.25in,clip,keepaspectratio]{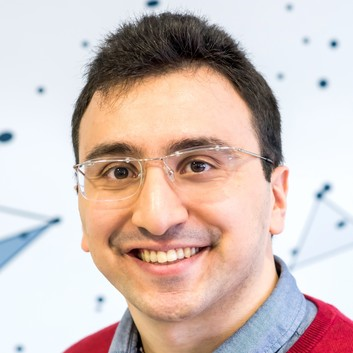}}]{Hany Assasa} 
is a senior system engineer at Pharrowtech, Belgium. He leads software and firmware activities for the Pharrowtech \SI{60}{\giga\hertz} RFIC module. Previously, he worked at IMDEA Networks Institute, Spain (2015 - 2020), first as a Ph.D. researcher and then as a Postdoctoral researcher. While working at IMDEA Networks, he focused on building efficient, robust, and reliable millimeter-wave wireless networks. His main interests are wireless networking, prototyping, wireless communications, and signal processing.
\end{IEEEbiography}

\begin{IEEEbiography}[{\includegraphics[width=1in,height=1.25in,clip,keepaspectratio]{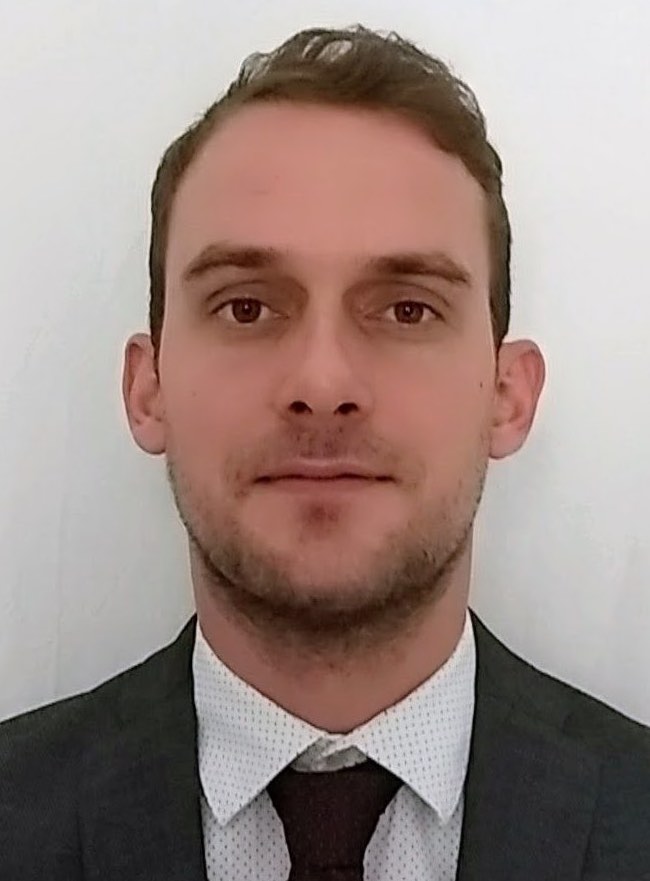}}]{Filip Lemic}
is a senior researcher at the i2Cat Foundation. He held positions at the University of Antwerp, imec, Universitat Politecnica de Catalunya, University of California at Berkeley, Shanghai Jiao Tong University, FIWARE Foundation, and Technische Universitat Berlin. He received his M.Sc. and Ph.D. from the University of Zagreb and Technische Universitat Berlin, respectively. 
\end{IEEEbiography}

\begin{IEEEbiography}[{\includegraphics[width=1in,height=1.25in,clip,keepaspectratio]{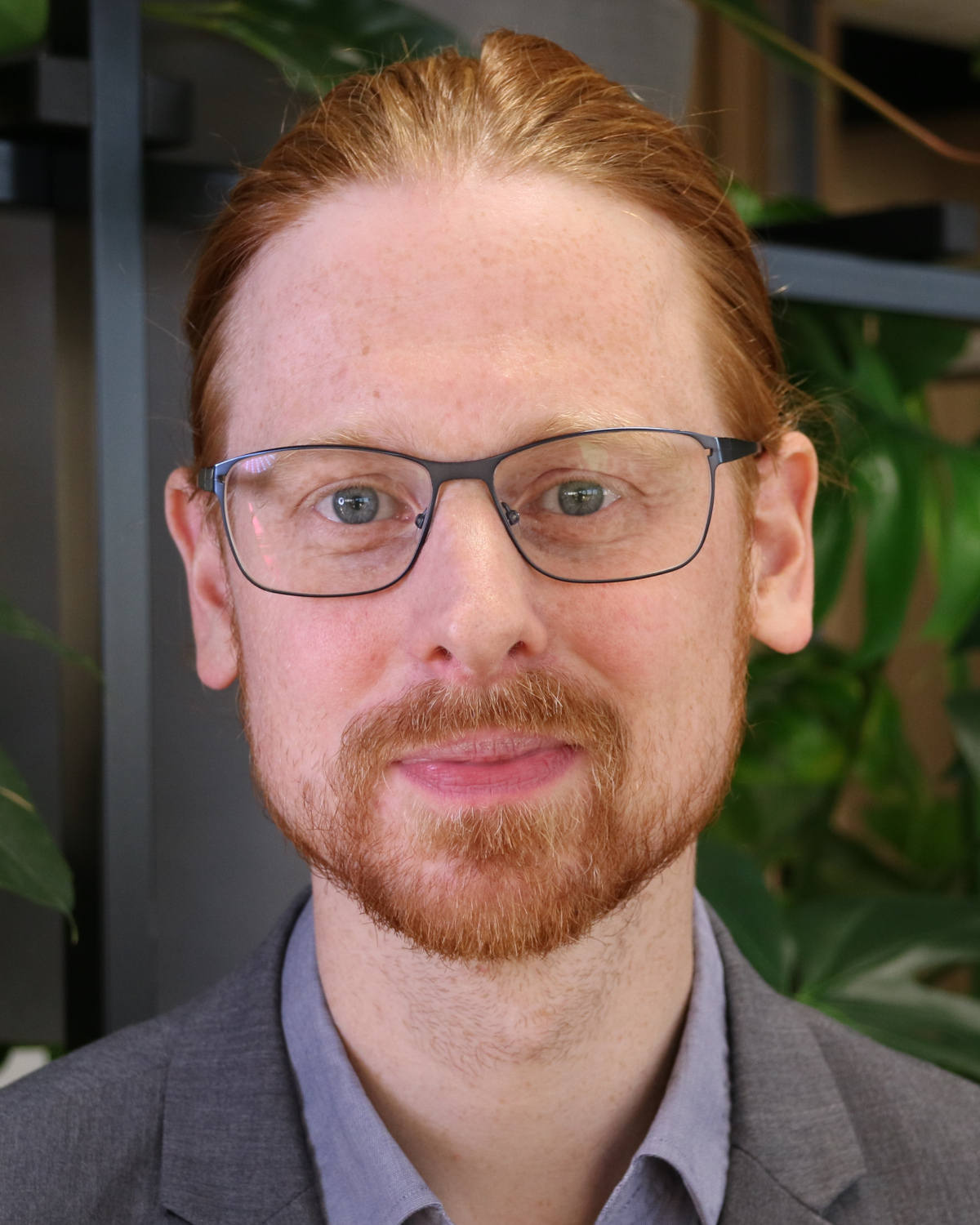}}]{Jeroen Famaey}
 is an associate research professor at the Department of Computer Science of the University of Antwerp and imec, Belgium. His research focuses on performance modeling and optimization of wireless network protocols. He has published over 170 peer-reviewed journal articles and conference papers, and 8 granted patents. He was listed as one of the top 2\% most cited scientists worldwide by Stanford University both in 2022 and 2023.
\end{IEEEbiography}

\begin{IEEEbiography}[{\includegraphics[width=1in,height=1.25in,clip,keepaspectratio]{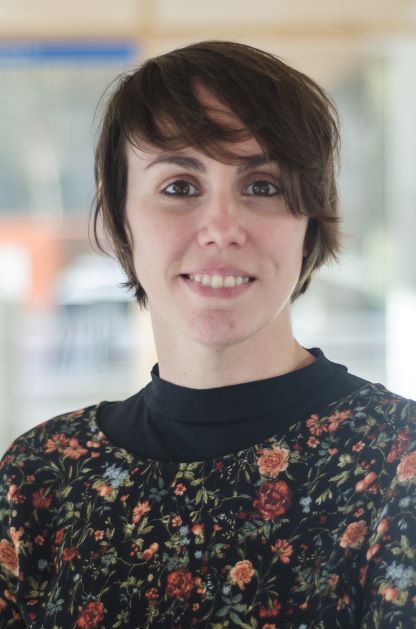}}]{Maria Torres Vega}
is a tenure track assistant professor at KU Leuven (Belgium), where her research focuses on devising human-driven control and management mechanisms for enhancing the perception of immersive systems. She received her M.Sc. degree in Telecommunication Engineering from the Polytechnic University of Madrid, Spain, in 2009 and her Ph.D. from the Eindhoven University of Technology, The Netherlands in 2017. Her research interests include quality of service and QoE in immersive multimedia systems and autonomous networks management.
\end{IEEEbiography}

\vfill

\end{document}